\documentstyle[12pt]{article}
\title{Determining the Penguin Effect on CP Violation in
$B^0 \rightarrow \pi^+ \pi^-$}
\author{Jo\~ao P.\ Silva and L.\ Wolfenstein \\
\small Department of Physics, Carnegie-Mellon University, \\
\small Pittsburgh, Pennsylvania 15213, U.S.A.}
\begin{document}
\maketitle
\begin{abstract}
A major goal in B physics is measuring the CP-violating asymmetry
in the decay $B^0 \rightarrow \pi^+ \pi^-$. In order to
determine one of the phase angles in the CKM matrix from this decay
it is necessary to determine the influence of the penguin amplitude.
Here we show how, using $SU(3)$ symmetry, the penguin effect can be
approximately determined from the ratio of the decay rates of
$B^0 \rightarrow K^+ \pi^-$ and
$B^0 \rightarrow \pi^+ \pi^-$.
\end{abstract}


Recently evidence has been presented for the decays
${\bar B}^0 \rightarrow \pi^+ \pi^-$ and
${\bar B}^0 \rightarrow \pi^+ K^-$ with a combined branching ratio of
$2 \times 10^{-5}$ \cite{exp}.
The decay ${\bar B}^0 \rightarrow \pi^+ \pi^-$ is of particular interest
since it is one of the prime candidates for the study of CP violation in
B decays.

{}From the sudy of CP violation in B decays, one can determine the phases
$\gamma$ and $\beta$ of the CKM matrix \cite{ckm},
defined by
\begin{eqnarray}
V_{ub} & = &
A \lambda^3 [\rho-i \eta] = A \lambda^3 z e^{-i \gamma}\ ,\\
V_{td} & = &
A \lambda^3 [(1-\rho)-i \eta] = A \lambda^3 y e^{-i \beta}\ ,
\end{eqnarray}
where
\begin{eqnarray}
y & = &
\sqrt {(1-\rho)^2 + \eta^2} =
(\cos{\gamma} + \sin{\gamma}/\tan{\beta})^{-1}\ ,
\label{eq:y}\\
z & = &
\sqrt{\rho^2 + \eta^2} = (\cos{\beta} + \sin{\beta}/\tan{\gamma})^{-1}\ .
\end{eqnarray}
The decay asymmetry for
$B^0 \rightarrow \Psi K_s$ determines the angle $\beta$ and,
in the tree approximation, the decay asymmetry for
$B^0 \rightarrow \pi^+ \pi^-$ determines $(\beta+\gamma)$.
However, a number of authors \cite{penguins} have emphasized
that there may be a sizeable uncertain penguin contribution to the
${\bar B}^0 \rightarrow \pi^+ \pi^-$ decay, thus making the determination
of $(\beta+\gamma)$ uncertain. Here we show that the approximate
$SU(3)$ symmetry can be used to estimate this penguin contribution once
the ratio of $\pi^+ K^-$ to $\pi^+ \pi^-$ is determined.

The effective hamiltonian responsible for these decays is \cite{buras}
\begin{equation}
H_{eff} = 2 \sqrt{2} G_F [\xi_u^\alpha (C_1 {\cal O}_1^\alpha
+ C_2 {\cal O}_2^\alpha) + \xi_t^\alpha \sum_{k=3}^6
 C_k {\cal O}_k^\alpha + h.c.]\ ,
\end{equation}
where
\begin{eqnarray}
\xi_u^\alpha & = &
V_{ub} V_{u \alpha}^\ast\ ,
\nonumber\\
\xi_t^\alpha & = &
V_{tb} V_{t \alpha}^\ast\ ,
\nonumber
\end{eqnarray}
and $\alpha = d$ for the $\pi^+ \pi^-$ decay and
$\alpha = s$ for the $\pi^+ K^-$ decay. The operators are
\begin{eqnarray}
{\cal O}_1^\alpha & = &
{\bar \alpha} \gamma^\mu \gamma_L b \ \
{\bar u} \gamma_\mu \gamma_L u\ ,
\nonumber\\
{\cal O}_2^\alpha & = &
{\bar u} \gamma^\mu \gamma_L b \ \
{\bar \alpha} \gamma_\mu \gamma_L u\ ,
\nonumber\\
{\cal O}_3^\alpha & = &
\sum_{q=u,d,s}\
{\bar \alpha} \gamma^\mu \gamma_L b \ \
{\bar q} \gamma_\mu \gamma_L q\ ,
\nonumber\\
{\cal O}_4^\alpha & = &
\sum_{q=u,d,s}\
{\bar q} \gamma^\mu \gamma_L b \ \
{\bar \alpha} \gamma_\mu \gamma_L q\ ,
\nonumber\\
{\cal O}_5^\alpha & = &
\sum_{q=u,d,s}\
{\bar \alpha} \gamma^\mu \gamma_L b \ \
{\bar q} \gamma_\mu \gamma_R q\ ,
\nonumber\\
{\cal O}_6^\alpha & = &
-2 \sum_{q=u,d,s}\
{\bar q} \gamma_L b \ \
{\bar \alpha} \gamma_R q\ ,
\end{eqnarray}
where $\gamma_{R,L} = (1 \pm \gamma_5)/2$.
The coefficients $C_k$ are calculated by the renormalization group equation.

The amplitude for the decays may be written as
\begin{eqnarray}
A({\bar B}^0 \rightarrow \pi^+ \pi^-) & = &
e^{-i \gamma} T_{\pi} + \lambda y e^{i \beta} P_\pi \ ,
\nonumber\\
A({\bar B}^0 \rightarrow \pi^+ K^-) & = &
- P_K + \lambda e^{-i \gamma} T_K \ ,
\end{eqnarray}
where
\begin{eqnarray}
T_\pi & = &
|V_{ub} V_{ud}|
<\pi^+ \pi^-| C_1 {\cal O}_1^d + C_2 {\cal O}_2^d|{\bar B}^0>\ ,
\nonumber\\
P_K & = &
|V_{tb} V_{ts}|
<\pi^+ K^-| \sum_{k=3}^6 C_k {\cal O}_k^s | {\bar B}^0>\ .
\end{eqnarray}
$T_K$ and $P_\pi$ are obtained by the substitutions
${\cal O}^d \rightleftharpoons {\cal O}^s$ and
$\pi^- \rightleftharpoons K^-$.
In the spectator approximation the decays are
\begin{eqnarray}
b ({\bar d}) & \rightarrow &
u ({\bar d}) + \pi^-\ ,
\nonumber\\
b ({\bar d}) & \rightarrow &
u ({\bar d}) + K^-
\end{eqnarray}

Since each ${\cal O}_k^d$ is related to ${\cal O}_k^s$ by an $SU(3)$
transformation ($180^o$ rotation around the V-spin axis),
it would follow in the
spectator and $SU(3)$ invariance approximations that
$T_K = T_\pi$ and $P_K = P_\pi$.
A formal evaluation of the $<{\cal O}^\alpha_k>$
matrix elements is typically done
using factorization. The results are quite uncertain, both because of the
uncertain $B \rightarrow \pi$ transition amplitude and the use of the
factorization approximation. We expect these two uncertanties to be
common to the two transitions so that we will use the factorization
result for the {\it ratio} of the matrix elements,
\begin{equation}
T_K/T_\pi = P_K/P_\pi = f_K/f_\pi\ .
\label{eq:formfact}
\end{equation}
In fact, the s and d quarks are much lighter than the b quark,
so that one expects the difference between the two to be negligable.
However, these
quarks will then hadronize into kaons and pions.
These hadronizations are different, as ilustrated by the difference between
their decay constants, $f_K$ and $f_\pi$.
Therefore it is natural to use Eq.~\ref{eq:formfact}
as an estimate of the SU(3) violation.

Neglecting final state interaction effects, $P$ and $T$ are real.
Then, the rate ratio is given by
\begin{eqnarray}
R & = &
\frac{\Gamma({\bar B}^0 \rightarrow \pi^+ K^-)}
{\Gamma({\bar B}^0 \rightarrow \pi^+ \pi^-)}
\nonumber\\
   & = &
(\frac{f_K}{f_\pi})^2 \frac{x^2 - 2 \lambda x \cos{\gamma} + \lambda^2}
{1 + 2 \lambda x y \cos{(\beta+\gamma)} + \lambda^2 y^2 x^2}\ ,
\label{eq:r}
\end{eqnarray}
where $x = P_K/T_K = P_\pi/T_\pi$ and $y$ is given by Eq.~\ref{eq:y}.
The asymmetry in the decay ${\bar B}^0 \rightarrow \pi^+ \pi^-$ is
given by
\begin{equation}
\alpha(\pi^+ \pi^-) = \sin{2(\beta +\gamma - \delta)}
\label{eq:asym}\ ,
\end{equation}
where $\delta$, the deviation of the phase of
$A({\bar B}^0 \rightarrow \pi^+ \pi^-)$ from its tree value,
is given by
\begin{equation}
\tan{\delta} = \frac{\lambda y x \sin{(\beta+\gamma)}}
{1+ \lambda y x \cos{(\beta+\gamma)}}
\label{eq:delta}
\end{equation}
Assuming the angle $\beta$ is determined from the asymmetry of the decay
into $\Psi K_s$ and that $R$ is measured, then for any value of
$\gamma$, we can use Eq.~\ref{eq:r} to determine $x$,
Eq.~\ref{eq:delta} to determine $\delta$, and
Eq.~\ref{eq:asym} to determine $\alpha(\pi^+ \pi^-)$.
There are two solutions for $x$;
in what follows we choose the solution with positive $x$,
as given by factorization.

Results are shown in Figs.~1 and 2. Figure 1, shows the allowed region
\cite{rhoeta} in the $(\rho,\eta)$ plane for $m_t < 180 GeV$.
Lines illustrating three fixed values of $\beta$ ($\beta=0.1 \pi/4$,
$0.25 \pi/4$ and $0.6 \pi/4$) are shown crossing this region.
While there is now no definitive data on $R$, the CLEO results
\cite{exp} favor a value between $1/3$ and $3$.
Therefore, we choose the values $R=1/3$, $1$ and $3$ in our illustrations.
Figs.~2 show the deviation of the asymmetry parameter
$\alpha(\pi^+ \pi^-)$ from its tree value, that is,
$\Delta \alpha(\pi^+ \pi^-) = \alpha(\pi^+ \pi^-) - \sin{2 (\beta + \gamma)}$,
as a function
of $\alpha(\pi^+ \pi^-)$, for our selected values of $\beta$ and $R$.

For small and moderate values of the magnitude of
$\alpha(\pi^+ \pi^-)$ there are significant deviations for values of
$R$ of order $1$ or greater. It is interesting to note that the size
of these deviations depends very little on the value of $\beta$.
For values of $\alpha(\pi^+ \pi^-)$ in the neighborhood of $+1$ ($-1$)
there are two solutions for $\gamma$ for fixed $\beta$,
even in the tree
approximation. These solutions correspond to values of
$(\beta + \gamma)$ either greater or smaller than
$\pi/4$ ($3 \pi/4$).
The sign of the deviation is seen to depend on which of the two solutions
is chosen.

We have ignored in these calculations final state interactions which
could produce a strong phase factor $\delta$ between the penguin and
tree terms. It is generally believed that $\delta$ is not very large.
The first order effect of $\delta$ is a term \cite{fsi}
proportional to $\sin{\delta}$ that shows up as a difference
in the decay rates of $B^0$ and ${\bar B}^0$. If we assume that
$R$ is calculated for the sum of the decays of
$B^0$ and ${\bar B}^0$, this effect cancels out.
Thus, the only effect of $\delta$ is proportional to
$(1-\cos{\delta})$ and should be unimportant for our results.

It has been pointed out by Gronau and others \cite{trapping}
that detailed studies of the time dependence of several $B^0$
decays could rigorously `trap' the penguin contribution.
Here we point out that, long before such measurements can be made,
one can obtain a reasonable approximate value for the penguin
effect simply from the measurement of branching ratios.


\vspace{5mm}

This work was supported by the United States Department of Energy,
under the contract DE-FG02-91ER-40682.
The work of J.\ P.\ S.\ was partially
supported by the Portuguese JNICT,
under CI\^{E}NCIA grant \# BD/374/90-RM.


%

\newpage


FIGURE CAPTIONS

\vspace{5mm}

Figure 1: Circular region defined by
the contraints on the CKM parameters
$\rho$ and $\eta$ from the combined measurements of $\epsilon$,
$|V_{ub}/V_{cb}|$ and $B^0 - {\bar B}^0$ mixing, for $m_t < 180 GeV$.
Also shown are presently allowed values of $\beta$ corresponding
to straight lines passing through $(\rho,\eta) = (1,0)$.
The solid line corresponds to $\beta = 0.6 \pi/4$, the dotted line to
$\beta = 0.1 \pi/4$. The dash-dotted corresponds to
$\beta = 0.25 \pi/4$ and crosses the allowed region twice.

Figures 2a,b,c: Curves of constant $\beta$ in the
$(\Delta \alpha(\pi^+ \pi^-),\alpha(\pi^+ \pi^-))$ plane,
for $R=1/3$ (Fig.~2a), $R=1$ (Fig.~2b) and $R=3$ (Fig.~2c).
The values of $\beta$ are those of Fig.\ 1.

\end{document}